\documentclass[12pt,a4paper,english,final]{article}
\usepackage[utf8]{inputenc}
\usepackage{float}
\usepackage{amsmath}
\usepackage{amssymb}
\usepackage{graphicx}
\makeatletter
\pdfpageheight\paperheight
\pdfpagewidth\paperwidth
\usepackage[english]{babel}
\usepackage[ final ]{pdfpages}
\DeclareMathOperator{\sign}{sign}
\usepackage{braket}
\usepackage{physics}
\usepackage{amsfonts}
\graphicspath{ {imagine/} }
\usepackage{subfigure}
\author{\textbf{Renzo Mosetti}\footnote{rmosetti@inogs.it}\\\ University of Trieste\\\
\small \textit{Department of Mathematics and Geoscience, Via Weiss, 4, 34127, Trieste, ITALY}}
\title{A non-local linear dynamical system and violation of Bell's inequality}

\usepackage{babel}

\makeatother

\usepackage{babel}
\begin{document}
\maketitle %\tableofcontents
%\newpage
\textbf{Abstract} A simple classical non-local dynamical system with
random initial conditions and an output projecting the state variable
on selected axes has been defined to mimic a two-channel quantum
coincidence experiment. Non-locality is introduced by a parameter connecting the initial conditions to the selection of the projection axes. The statistics of the results shows violations up to 100$\%$ of the Bell's inequality, in the form of Clauser-Horne-Shimony-Holt (CHSH), strongly depending on the non-locality parameter. Discussions on the parallelism with Bohmian mechanics are given. \newpage{}

\section{Introduction}

The Bell inequality \cite{bell} \cite{bell1} is a milestone in the
development of the post EPR \cite{einstein} discussion on the completeness
of Quantum Mechanics. The literature on the subject is huge and laboratory
experiments confirm the violation of the inequality in the quantum
world. Moreover, there are also several papers on the occurrence of
Bell's inequalities violation in classical random systems \cite{aerts},
\cite{aerts1}, \cite{kofler}, \cite{vangehr}, \cite{clemente}
. In this contest, the Accardi et al. papers, \cite{accardi}, \cite{accardi1}
have raised an animate debate due mainly by Gill ( \cite{gill}, \cite{gill1}
). Usually, the experiments are based on computer randomness generation
while in this paper we perform correlation experiments much possible
close to the concept of the evolution of a dynamical state which resembles
the causal Bohmian interpretation of EPR where initial conditions
of the wave function appear to be crucial. In fact, the complete Bohm
theory \cite{bohm} is based on a deterministic dynamics and one could
introduce randomness \textit{by randomizing the initial conditions} according to the well known Born's rule.
So, the randomness of the quantum theory is due to our ignorance of
initial conditions. The numerical experiment here proposed lies within
these lines. In the first paragraph a simple two dimensional state-space
dynamical system with a scalar output is defined. In the second paragraph
we define the experiment which mimics the behavior of the quantum
two-channel polarized photon experiment for testing the Clauser-Horne-Shimony-Holt
(CHSH) inequality \cite{clauser} . The third paragraph contains the
set-up of the simulator and algorithm used to compute correlation
among measurement pairs. The fourth paragraph presents the obtained
results in terms of the statistics of the outputs. Finally, some conclusions
and open questions are given.

\section{The state-space linear system model}

Consider the following two-dimensional linear dynamical system in
the state-space representation: \\
 \
\begin{equation}
\begin{aligned}\dot{X}=AX\\
Y=CX
\end{aligned}
\end{equation}
$A=\begin{pmatrix}0 & 1\\
-1 & 0
\end{pmatrix}$; $\hspace{1cm}C=\begin{bmatrix}\cos(\theta) & \sin(\theta)\end{bmatrix}$;
$\hspace{1cm}X^{T}:=\begin{bmatrix}x_{1} & x_{2}\end{bmatrix}$ \\
 \\
 The formal solution for the state propagation starting with an initial
condition: $X_{0}:=X(t=0)$ is: 
\begin{equation}
X(t)=X_{0}e^{tA}
\end{equation}
\\
 \
In this case, the solution is simply a sinusoidal wave. The output
is the projection of the state vector along the axis defined by $\theta$:
\begin{equation}
y(t)=\begin{bmatrix}\cos(\theta) & \sin(\theta)\end{bmatrix}\begin{bmatrix}x_{1}(t)\\
x_{2}(t)
\end{bmatrix}
\end{equation}

\section{The design of the experiment }

The usual Alice and Bob perform separate measurements from the output
$y$ of the system (1) at time $t=t_{1}$ by choosing independently
an angle $\theta$. \\
 The scheme is reported in Figure 1. 
\begin{figure}[H]
\centering \includegraphics[scale=0.47]{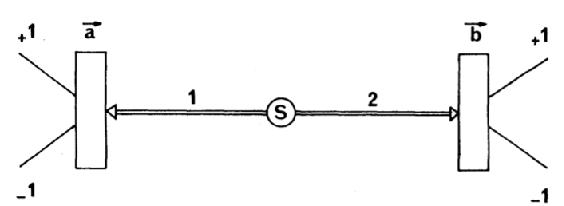} \label{label}
\caption{Scheme of the experiment: S can be considered as a source as well
as the black box containing the system state}
\end{figure}

The results coming from Alice measurements are labeled with $\lbrace a\rbrace$
and those from Bob with $\lbrace b\rbrace$ . \\
 In order to reproduce the typical measurements of a quantum system,
we define the measurements in the set $\lbrace-1,+1\rbrace$ according
to the following rule:\
\begin{equation}
\begin{aligned}a_{i}=\sign[y_{a}(t=t_{1})](i=1,2)\\
b_{i}=\sign[y_{b}(t=t_{1})](i=1,2)
\end{aligned}
\end{equation}
\\
 The idea is to test, by means of a series of experiments, whether
it is possible to violate the Bell's inequality by testing the experimental
correlations defined in a standard way as: 
\begin{equation}
E=(N_{++}+N_{--}-N_{+-}-N_{-+})/(N_{++}+N_{--}+N_{+-}+N_{-+})
\end{equation}
Where: $N_{++}$ means that $(a=+1)\wedge(b=+1)$ and so on. Actually,
being the experiment a typical two-channel experiment, the CHSH inequality
in the following form is more suitable: 
\begin{equation}
S=E(a_{1}b_{1})-E(a_{1}b_{2})+E(a_{2}b_{1})+E(a_{2}b_{2})\leqslant2
\end{equation}
Furthermore, the derivation of the original Bell inequality assumes
perfect anticorrelation of entangled photon measurements in the EPR
setup. The CHSH inequality has the reputation of being superior for
all purposes. It is harder to violate and its derivation \textit{does
not need anti-correlation}. The following hypotheses are assumed to
derive the CHSH inequality :\\
 \\
 \textit{Hidden variables:} The results of any measurement on any
individual system are predetermined;\\
 \\
 \textit{Non}-\textit{Locality}: It is well known that non-locality is necessary to reproduce the quantum mechanics in a deterministic context. So, non-local effect are here introduced in the initial conditions (hidden variables) by the following rule:

\[
x_{1}(0)=r_{1}+\lambda\cos(\vartheta_{1});
\]

\[
x_{2}(0)=r_{2}+\lambda\cos(\vartheta_{2});
\]
\\
where: $r_{1}$ and $r_{2}$ are Gaussian random noises with $0$ mean and unitary variance;
and $\lambda\cos(\vartheta_{1})$ and $\lambda\cos(\vartheta_{2})$ constitute 
the dependence of the initial conditions on the chosen angle by both  Bob and Alice experimental setting.
The violation of a Bell inequality
guarantees that the observed outputs are not predetermined and
that they arise from entangled quantum systems that possess intrinsic
randomness.
As for the standard tests for quantum systems, the experiments consist
on four sub-experiments (each for any correlation pair) and the state
model is \textit{the same both for Alice and Bob} who do not know
the system state and perform the measurements separately at the same
time $t_{1}$. To randomize the system, the initial condition of the
state variable $X_{0}$ is picked-up from a normal distribution with
zero mean and unitary variance according to the non-local position
described in the formula above in \textit{every run of each sub-experiment}.
To obtain a significant statistics, a large number of experiments
have to be performed by iterating the procedure. In an algorithmic
language, the process can be synthesized as follows: \\
 \\
 Set $\lambda$ \\
 \
 For n=1,Nmax \\
 \
 For i=1,M \\
 \
Do {[}random $X_{0}$, State propagation$\Rightarrow$ measurement
$a_{1}$ and $b_{1}${]} \\
 \
 end \\
 \
Compute correlation according to (5)$E(a_{1}b_{1})$ \\
 \ For i=1,M \\
 \
Do {[}random $X_{0}$, State propagation$\Rightarrow$ measurement
$a_{1}$ and $b_{2}${]} \\
 \
 end \\
 \
Compute correlation $E(a_{1}b_{2})$ \\
 \ For i=1,M \\
 \
Do {[}random $X_{0}$, State propagation$\Rightarrow$ measurement
$a_{2}$ and $b_{1}${]} \\
 \
 end \\
 \
Compute correlation $E(a_{2}b_{1})$ \\
 \ For i=1,M \\
 \
Do {[}random $X_{0}$, State propagation$\Rightarrow$ measurement
$a_{2}$ and $b_{2}${]} \\
 \
 end \\
 \
Compute correlation $E(a_{2}b_{2})$ \\
 \ Compute S(n) \\
 \
 end \\
 \ Compute statistics on S \\
\section{Running the model}

We chose as angles for the coincidence measurements the ones which,
according to the quantum theory, give $S=2\sqrt{2}$. 
\begin{equation}
\lbrace\theta\rbrace=\lbrace\dfrac{\pi}{2},0,\dfrac{\pi}{4},-\dfrac{\pi}{4}\rbrace
\end{equation}
The importance of this configuration lies on the fact that it corresponds
to a situation which \textit{cannot be reproduced by a hidden variable
theory}. We have performed 100 (Nmax) independent runs with M=100
yielding hundred measurement pairs in each sub-experiment. The $Mathematica^{TM}$
software has been used for the simulations and for the statistical
analyses of the outputs. In what follows, the measurement pairs with
respect to the chosen angles are reported in details.

\paragraph{First coincidence measurements}

\begin{equation}
\begin{aligned}a_{1}=\sign[y_{a}(t_{1})]=\begin{bmatrix}\cos(\dfrac{\pi}{2}) & \sin(\dfrac{\pi}{2})\end{bmatrix}\begin{bmatrix}x_{1}(t_{1})\\
x_{2}(t_{1})
\end{bmatrix}\\
b_{1}=\sign[y_{b}(t_{1})]=\begin{bmatrix}\cos(\dfrac{\pi}{4}) & \sin(\dfrac{\pi}{4})\end{bmatrix}\begin{bmatrix}x_{1}(t_{1})\\
x_{2}(t_{1})
\end{bmatrix}
\end{aligned}
\end{equation}

\paragraph{Second coincidence measurements}

\begin{equation}
\begin{aligned}a_{1}=\sign[y_{a}(t_{1})]=\begin{bmatrix}\cos(\dfrac{\pi}{2}) & \sin(\dfrac{\pi}{2})\end{bmatrix}\begin{bmatrix}x_{1}(t_{1})\\
x_{2}(t_{1})
\end{bmatrix}\\
b_{2}=\sign[y_{b}(t_{1})]=\begin{bmatrix}\cos(-\dfrac{\pi}{4}) & \sin(-\dfrac{\pi}{4})\end{bmatrix}\begin{bmatrix}x_{1}(t_{1})\\
x_{2}(t_{1})
\end{bmatrix}
\end{aligned}
\end{equation}

\paragraph{Third coincidence measurements}

\begin{equation}
\begin{aligned}a_{2}=\sign[y_{a}(t_{1})]=\begin{bmatrix}\cos(0) & \sin(0)\end{bmatrix}\begin{bmatrix}x_{1}(t_{1})\\
x_{2}(t_{1})
\end{bmatrix}\\
b_{1}=\sign[y_{b}(t_{1})]=\begin{bmatrix}\cos(\dfrac{\pi}{4}) & \sin(\dfrac{\pi}{4})\end{bmatrix}\begin{bmatrix}x_{1}(t_{1})\\
x_{2}(t_{1})
\end{bmatrix}
\end{aligned}
\end{equation}

\paragraph{Fourth coincidence measurements}

\begin{equation}
\begin{aligned}a_{2}=\sign[y_{a}(t_{1})]=\begin{bmatrix}\cos(0) & \sin(0)\end{bmatrix}\begin{bmatrix}x_{1}(t_{1})\\
x_{2}(t_{1})
\end{bmatrix}\\
b_{2}=\sign[y_{b}(t_{1})]=\begin{bmatrix}\cos(-\dfrac{\pi}{4}) & \sin(-\dfrac{\pi}{4})\end{bmatrix}\begin{bmatrix}x_{1}(t_{1})\\
x_{2}(t_{1})
\end{bmatrix}
\end{aligned}
\end{equation}

\subsection{Results: CHSH inequality violation}

For each selected values of $0\leq\lambda \leq 10$  the process, as described above, has been run and the results obtained  of the overall expected value of $S$ and on the percentage of CHSH violations and are reported in the Figures 2 and 3 respectively.  By performing several runs, also by increasing the number M of measurements pairs and the number of
iterations, the overall statistics do not show any significant variation.

\begin{figure}[H]
\centering \includegraphics[scale=1]{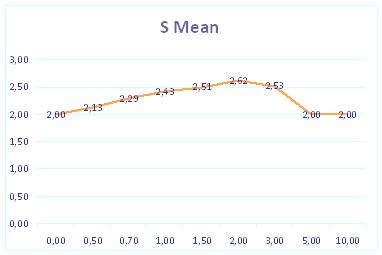}
\label{label} \caption{S mean values as a function of $\lambda$} 
\end{figure}

\begin{figure}[H]
\centering \includegraphics[scale=1]{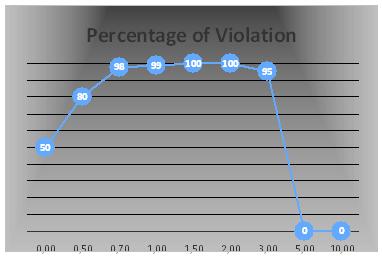}
\label{label} \caption{Percentage of violation as a function of $\lambda$} 
\end{figure}

Furthermore, the following figures show the statistics  of the run with $\lambda=2$ corresponding to the maximum value of the average of $S$ i.e. $\langle S\rangle=2.62$ with a standard
deviation $STD(S)=0.12$. Note that this value is not to far from the quantum
value $S=2\sqrt{2}\simeq 2.83$.

\begin{figure}[H]
\centering \includegraphics[scale=0.5]{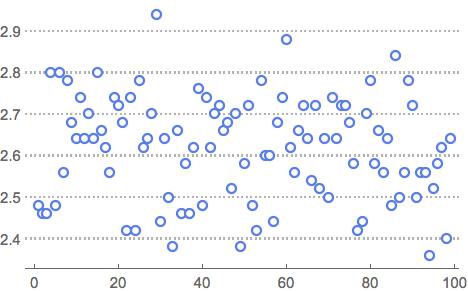}
\label{label} \caption{Iteration number and corresponding S value ($\lambda=2$)}
\end{figure}

\begin{figure}[H]
\centering \includegraphics[scale=0.5]{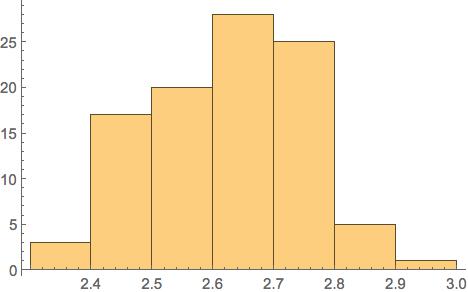}
\label{label} \caption{Histogram of the values of S violating the CHSH inequality ($\lambda=2$)}
\end{figure}

\begin{figure}[H]
\centering \includegraphics[scale=0.5]{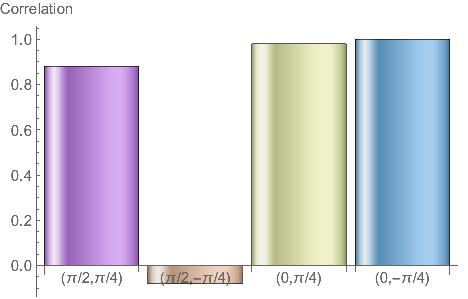}
\label{label} \caption{Correlation values of the maximal violation ($S=2.94$) of CHSH inequality ($\lambda=2$)}
\end{figure}

If we look to Fig. 2 and 3, we notice that for an experiment forced to locality i.e. $\lambda=0$, the expectation value is 2, and  50$\%$ of the runs  violate the inequality. This is expected from the completely random nature of the output (Fig. 7). As the effect of non-locality affects the initial condition the violation is growing up to the maximum value for $S$ mean corresponding to $\lambda=2$. By further increasing $\lambda$ a decay on violation up to 0 percentage and again an average value of $S$  equal to 2 is obtained. This because the randomness of initial conditions are negligible and the system becomes a causal deterministic system yielding the upper limit of the CHSH inequality.
 
\begin{figure}[H]
\centering \includegraphics[scale=0.5]{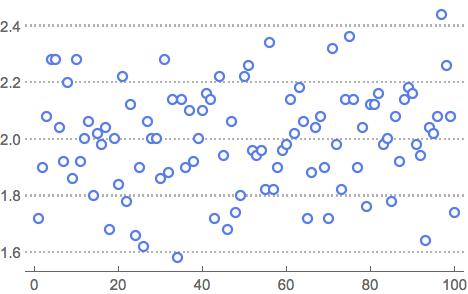}
\label{label} \caption{Iteration number and corresponding S value for $\lambda=0$}
\end{figure}

\section{Conclusions}

We do not want to force any \char`\"{}philosophical\char`\"{} speculation
on the results. It is well known that an impenetrable wall separating
Quantum from Classical Mechanics does exist. We simply enumerate the
following facts.\\
 Coincidence measurements from a deterministic but non-local and simple linear dynamical
model with random initial conditions show violations of the CHSH inequality
up to hundred percent of coincidence measurements and the
statistics of the results are robust against the large number of repeated
experiments and the number of the measurement values. In the case of locality the results are in agreement with other random simulations. As Gill pointed out,: \char`\"{}it is easy to make simulation models of local realistic loophole free CHSH-type experiments which violate
CHSH i.e. the mean value of S is 2 and half the time the experiment
gives a larger result and half the time it gives a smaller result\char`\"{}
(https://pubpeer.com/publications
/B087561066AD645C5348ADC2E4CF1C).
The approach of our experiment is to mimic through a simple dynamical
system some aspects of the Bohmian mechanics. Bohmian mechanics is a deterministic
theory where a probabilistic element is introduced as probability
is introduced into classical mechanics. The observer does not possess
the full information about the true initial configuration of the system,
but rather he has to retreat to a probability density. (See for example:
\cite{kim} for a criticism of Bohmian theory). Of course, we cannot
conclude that our results might imply adequacy or inadequacy of Bohmian point of
view.

\end{document}